\pdfoutput=1

\documentclass[
  twocolumn,
  prb,
  amsmath,
  amssymb,
  superscriptaddress
]{revtex4}
\usepackage{hyperref}

\usepackage{notes2bib}
\bibnotesetup{note-name    = ,use-sort-key = false}

\usepackage{bm}
\usepackage{graphicx}
\usepackage{amssymb}
\usepackage{color}
\usepackage{braket}
\usepackage{pifont}
\usepackage{empheq}
\usepackage{scalerel,stackengine}
\stackMath
\newcommand\reallywidehat[1]{%
\savestack{\tmpbox}{\stretchto{%
  \scaleto{%
    \scalerel*[\widthof{\ensuremath{#1}}]{\kern-.6pt\bigwedge\kern-.6pt}%
    {\rule[-\textheight/2]{1ex}{\textheight}}
  }{\textheight}%
}{0.5ex}}%
\stackon[1pt]{#1}{\tmpbox}%
}

\renewcommand{\v}[1]{\textbf{\textit #1}}

\newcommand{\blue}[1]{\textcolor{blue}{#1}}

\begin{document}

\title{Resistance of 2D superconducting films}

\author{E.~J.\ K\"onig}
\affiliation{Max Planck Institute for Solid State Research, D-70569 Stuttgart, Germany}

\author{I.~V.\ Protopopov}
\affiliation{Department of Theoretical Physics, University of Geneva, 1211 Geneva, Switzerland}
 \affiliation{L.~D.\ Landau Institute for Theoretical Physics RAS, 119334 Moscow, Russia}

\author {A.\ Levchenko} 
\affiliation{Department of Physics, University of Wisconsin-Madison, Madison, Wisconsin 53706, USA}

\author{I.~V.\ Gornyi}
\affiliation{\mbox{Institute for Quantum Materials and Technologies, Karlsruhe Institute of Technology, 76021 Karlsruhe, Germany}}
\affiliation{\mbox{Institut f\"ur Theorie der kondensierten Materie, Karlsruhe Institute of Technology, 76128 Karlsruhe, Germany}}
 \affiliation{Ioffe Institute, 194021 
 St.~Petersburg, Russia}

  \author{A.~D.\ Mirlin}
\affiliation{\mbox{Institute for Quantum Materials and Technologies, Karlsruhe Institute of Technology, 76021 Karlsruhe, Germany}}
\affiliation{\mbox{Institut f\"ur Theorie der kondensierten Materie, Karlsruhe Institute of Technology, 76128 Karlsruhe, Germany}}
 \affiliation{L.~D.\ Landau Institute for Theoretical Physics RAS, 119334 Moscow, Russia}
  \affiliation{Petersburg Nuclear Physics Institute, 188300 St.~Petersburg, Russia. }

\begin{abstract}
{We consider the problem of finite resistance $R$ in superconducting films with geometry of a strip of width $W$ near zero temperature}. The resistance is generated by vortex configurations of the phase field. In the first type of process, \textit{quantum phase slip}, the vortex worldline in 2+1 dimensional space-time is space-like (i.e., the superconducting phase winds in time and space). In the second type, 
\textit{vortex tunneling}, the worldline is time-like (i.e., the phase winds in the two spatial directions) and connects opposite edges of the film.
For moderately disordered samples, processes of second type favor a train of vortices, each of which tunnels only across a fraction of the sample. Optimization with respect to the number of vortices yields a tunneling distance of the order of the coherence length $\xi$, and the train of vortices becomes equivalent to a quantum phase slip. Based on this theory, we find the resistance $\ln R \sim -g W/\xi$, where $g$ is the dimensionless normal-state conductance. Incorporation of quantum fluctuations indicates a quantum phase transition to an insulating state for $g \lesssim 1$.
\end{abstract}


\maketitle

\textit{\blue{Introduction.}} Since its {original discovery by H. Kamerlingh Onnes} 
a defining feature of superconductivity is its vanishing electrical resistivity in the thermodynamic limit. Yet, experimental samples are finite and can therefore be expected to display a non-zero, albeit small, resistance $R$. This naturally provokes questions about the resistance of superconductors~\cite{Huse1992,HalperinRefaelDemler2011} 
-- particularly in two-dimensional (2D) films, where resistivity and resistance have the same physical dimension: 
What is the dependence of $R$ on the system size and disorder strength? Is its scaling with increasing system size ``dual'' to the exponentially vanishing zero-temperature conductance of an insulator? 

\begin{figure}
\includegraphics[width = .48\textwidth]{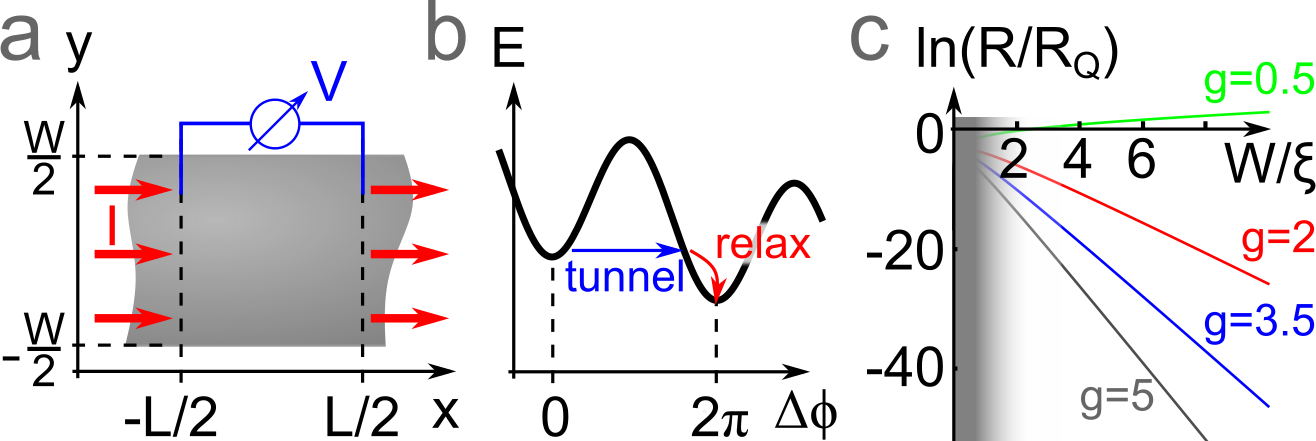} 
\caption{(a) Setup for a prototypical measurement of resistance in a finite 2D superconducting film. (b) The bias current $I$ lifts the periodicity of the ground state energy as a function of a fixed $\Delta \phi = \phi(L/2)-\phi(-L/2)$. Tunneling events (and subsequent energy relaxation) between adjacent minima generate resistance. (c) Resistance, Eq.~\eqref{eq:LogRFinal}, per square  (i.e., $RW/L$) for several values of the normal-state conductance $g$. The linearly decaying asymptotes of $\ln R$ are consistent with experiment.~\cite{SchneiderStrunk2019}}
\label{fig:Setup}
\end{figure}

An approximate duality of this kind can be formalized~\cite{Fisher1990} using the particle-vortex duality within the bosonic description of the superconductor-insulator 
transition  (SIT),~\cite{Gantmakher2010,LinGoldman2015,SacepeKlapwijk2020} when the normal state resistance is tuned to the quantum of resistance $R_Q \equiv h/4e^2$. 
%
Such theories are optimized for samples, where
Cooper pairs on superconducting granules Anderson-delocalize at the SIT. In this paper, instead, we concentrate on the different experimental situation of homogeneous films for which the impurity induced reduction and ultimate annulment of $T_c$ (as defined by the onset of a spectral gap) is well described by fermionic theories
~\cite{Finkelstein1990,BurmistrovMirlin2015} and 
the study of resistivity below $T_c$ constitutes a separate, subsequent question: At finite temperature, but infinite system size, resistance is established by vortex proliferation above a renormalized Berezinskii-Kosterlitz-Thouless temperature.~\cite{HalperinNelson1979,KoenigMirlin2015}  
On the other hand, near zero temperature, but at finite system size, the
%
resistance 
as a function of size and disorder strength is unknown and is the subject of this paper. 
We focus on a 2D superconducting strip of width $W$ and length $L > W$, Fig.~\ref{fig:Setup}a.

We begin by briefly reviewing the literature on the resistance of superconductors. Josephson's equation relates the voltage
$V =  ({d \Delta \phi}/{d t}){\hbar}/2 e$
to the time derivative of the phase difference ${\Delta\phi=\phi(x = L/2)-\phi(x = -L/2)}$. The presence of a bias current $I$ lifts the degeneracy of quantum states with integer difference in $\Delta \phi/2\pi$, see Fig.~\ref{fig:Setup}b. Near zero temperature, quantum tunneling between adjacent minima dominates the decay of the phase difference. The voltage is then given by the effective tunneling rates $1/\tau_{\pm 2\pi}$ for the change of the phase by $\pm 2\pi$, 
\begin{equation}
V = \frac{h}{2 e}\left(\frac{1}{\tau_{2\pi}}- \frac{1}{\tau_{-2\pi}}\right). \label{eq:VoltageDecayrate}
\end{equation}
Theoretically, the decay of metastable vacua is described by instanton-like field configurations \cite{Vainshtein1964,Langer1967,Coleman1977,CallenColeman1977} i.e., imaginary-time solutions of the semiclassical equations of motion which connect the adjacent minima. For the decay of the supercurrent, the general form of such solutions is not known. However, certain trial solutions given by vortex configurations are believed to be good approximations. For the present strip geometry, those are the following: (i) vortex configurations which are $y$-independent and swirl in the $x-\tau$ plane ($\tau$ is the imaginary time), i.e., the worldline of the core is space-like; (ii) vortices which wind in the $x-y$ plane and tunnel across the system perpendicularly to the current (a time-like worldline).
We will refer to case (i)  as \textit{quantum phase slips} and (ii) as \textit{vortex tunneling}. 

Most of the literature on the resistance of superconductors is devoted to thermally activated decay of the supercurrent~\cite{LangerAmbegaokar1967,McCumberHalperin1970,OvchinnikovVarlamov2015}
and to vortex physics in the presence of a magnetic field~\cite{Anderson1962,BlatterGeshkenbeinVinokur1991,BlatterRMP1994} or for current bias close to the critical current. We restrict ourselves to summarizing the work in the quantum regime for quasi-1D and 2D systems at small current and without external field.
Supercurrent decay in homogeneous 1D wires was considered theoretically in Refs.~[\onlinecite{SaitoMurayama1989,Duan1995,RennDuan1996,ZaikinZimanyi1997,GolubevZaikin2001,Khlebnikov2004,RefaelFisher2007}]. There is a consensus that the semiclassically dominant field configuration is a dipole of phase slips at distance $\delta \tau \propto 1/I$ in time direction, while the role of electromagnetic fields led to a debate~\cite{Duan1997,ZaikinZimanyi1997Reply} between the authors of Refs.~[\onlinecite{Duan1995}] and [\onlinecite{ZaikinZimanyi1997}]. In addition, it was shown that inhomogeneities in the wire can be crucially important.~\cite{PaiAndrei2008,VanevicNazarov2012}
For 2D samples, studies of voltage generation in ultraclean samples focus on the theory of vortex tunneling,\cite{AoThouless1994,IengoJug1995,IengoJug1996,ArovasAuerbach2008} partly in terms of a very elegant mapping~\cite{Popov1973,LeeFisher1991} to 2+1 dimensional electrodynamics.
At the same time,  the role of (i) finite width (ii) electromagnetic gauge potentials were disregarded in these works. Further, the influence of the Magnus force and the size of the vortex mass remained controversial (for review, see Ref.~[\onlinecite{ThoulessWexler1999}]). Quantum tunneling in the presence of inhomogeneities (pinning centers) was addressed in Refs.~[\onlinecite{FeigelmanLevit1993,AoThouless1994,Stephen1994,AuerbachGhosh2006}]. 

Experimental evidence \cite{Tafuri2006} for quantum tunneling of vortices in 2D superconductors at finite current bias, in particular in the context of dark photon  counts,\cite{Kitaygorsky2007,Bulaevskii2011} additionally augmented the interest in this research field. New experimental tools, such as SQUID-on-tip microscopy,\cite{EmbonZeldov2017} allow accessing both vortex motion and energy dissipation (local heating). At the same time, the direct measurement of the extremely small resistance of large samples of 2D superconductors is experimentally challenging; most recent studies~\cite{TamirShahar2019} demonstrate the technical subtleties and advances in the careful filtering of external radiation. A remedy is to study higher-resistance samples, e.g., not too far from the SIT or with a smaller width $W$. In the regime where a finite zero-temperature saturation value of $R$ is measurable, experimental data~\cite{SchneiderStrunk2019} is consistent with $- \ln R \propto  W$.

In this work, we consider moderately disordered homogeneous films with $1/\tau_{\rm el} \gg \Delta^2/E_F$ under infinitesimal current bias. We will refer to the case $1/\tau_{\rm el} \ll \Delta$ ($1/\tau_{\rm el} \gg \Delta$) as clean (dirty) limit. Here, $E_F$ is the Fermi energy, $\tau_{\rm el}$ the elastic scattering time of electrons, and $\Delta$ the spectral gap of the superconductor. As discussed below, for systems considered here the Magnus force is unimportant and the vortex mass is finite. We find that the combined tunneling of several vortices dominates over single-vortex events and demonstrate that, for the optimal number of involved vortices, the tunneling process has the same contribution as quantum phase slips. This allows us to determine the linear-response resistance of superconducting strips, which is exponentially small but finite.
Estimating the preexponential factor and pushing the theory to the border of its applicability range, we also show that the SIT due to the antagonistic interplay of energy versus entropy (here due to quantum fluctuations) is captured at sufficiently strong disorder $1/\tau_{\rm el} \sim E_F$.

\textit{\blue{Quantum phase slips.}} We first consider the voltage generation by quantum phase slips. 
In perfectly coherent systems, the decay out of a ``false'' vacuum is technically encoded in ``bounce solutions'':~\cite{Coleman1977,ZaikinZimanyi1997} a phase slip takes the system to a lower vacuum where it dwells for a time $\delta \tau$ and subsequently an anti-phase-slip takes it back to the point of departure. The action of a bounce is ($\Phi_0$ is the flux quantum)
\begin{equation}
S_{I}(\delta \tau) \simeq 2 S_{\rm core} + K_{\rm 1D} \ln(\Delta \delta \tau) - \Phi_0 I \delta \tau. 
\label{eq:Bounce}
\end{equation}
Each phase slip described by Eq.~\eqref{eq:Bounce} is weighted by its core action $S_{\rm core} \sim K_{\rm 1D}\sim (W/\xi) E_F \min(\tau_{\rm el}, 1/\Delta)$, where the factor $(W/\xi)$ represents the length of the space-like worldline of the core. Here, we introduced the dimensionless stiffness of 1+1 dimensional (i.e., $y$-independent) phase fluctuations $K_{\rm 1D}$ and the coherence length $\xi \sim v_F/\Delta$ ($\xi \sim v_F\sqrt{\tau_{\rm el}/\Delta}$) in the clean (dirty) limit, where $v_F$ is the Fermi velocity. 

The action is maximal at the typical dwell time $\delta \tau_{\rm typ} = K_{\rm 1D}/(\Phi_0 I)$, and a steepest-descent evaluation~\bibnote
{Reference~\onlinecite{ZaikinZimanyi1997} evaluated the integral $\int d \delta \tau \exp[-S(\delta \tau)]$ exactly and found an exponent $R\sim I^{K_{\rm 1D} - 2}$. It is in agreement with the result $R\sim I^{K_{\rm 1D} -1}$ of the steepest-descent method used in our work in its regime of applicability, $K_{\rm 1D} \gg 1$. The advantage of our approach is that it is applicable for both quantum phase slips and 2D vortex tunneling.} 
of the instanton contribution to the partition sum \cite{Kleinert2009} $\mathcal Z_{\rm instanton} \sim (L/\xi)\int d \delta \tau \exp[-S_I(\delta \tau)]$ 
yields the decay rate~\cite{ZaikinZimanyi1997} $1/\tau_{2\pi} \sim \Delta^2 \delta \tau_{\rm typ}\exp[{- S_I(\delta \tau_{\rm typ})}]L/[\xi \sqrt{K_{\rm 1D}}] \sim I^{K_{\rm 1D} - 1} $. This non-linear current-voltage characteristics is a manifestation of a perfectly quantum-coherent bounce. However, at infinitesimal bias current, $\delta \tau_{\rm typ}$ exceeds the relaxation time $\tau_{\rm coh}$ which is always finite in a finite system connected to the metallic leads or in the presence of an external bath. The broadening of levels inside the core described by $\tau_{\rm coh}$ leads to a finite rate for quantum phase slips and anti-phase-slips
${1}/{\tau_{\pm 2\pi}} \propto  e^{-S_{\pm I}(\tau_{\rm coh})}L/(\tau_{\rm coh} \xi).$
Then, Eq.~\eqref{eq:VoltageDecayrate} leads to a linear current-voltage relation,
\begin{equation}\label{eq:RSlips}
R_{\rm {QPS}} \sim \frac{h}{e^2} \frac{L}{\xi} K_{\rm 1D}^{-3/2} \left (\tau_{\rm coh} \Delta\right)^{2-K_{\rm 1D}}e^{-2 K_{\rm 1D}},
\end{equation}
where the prefactor is determined by matching the non-linear resistance at $I \gtrsim K_{\rm 1D}/(\Phi_0 \tau_{\rm coh})$ mentioned above. 
 
We conclude this consideration with three remarks. First, we comment on $\tau_{\rm coh}$ which is finite in any realistic situation and may result from a variety of system-dependent origins, such as phonons, external radiation, and noise in the leads. Different sources generally imply different temperature and system-size dependence of the relaxation time which we do not study here, since the main factor in $-\ln R$ is the tunneling action, while $\tau_{\rm coh}$ enters in the form of a logarithmic prefactor. 
Second, spatial fluctuation of the worldline of the  phase-slip core around the $y-$independent line have been disregarded up to now. These fluctuations lead to an additional preexponential factor, which we estimate at the end of the paper. Third, magnetic screening effects (see 
Ref.~[\onlinecite{Duan1997}] and Supplement~\cite{Supplement}) are unimportant in the above consideration~\cite{ZaikinZimanyi1997Reply} so long as $\tau_{\rm coh} \Delta  \ll \exp(\lambda_M/W)$, where $\lambda_M$ is the Pearl length  (2D analog of the London penetration length), which is usually macroscopic. For longer relaxation times, the result for $-\ln R$ gets modified according to $K_{\rm 1D} \ln(\tau_{\rm coh} \Delta ) \rightarrow K_{\rm 1D} \lambda_M\ln[W\ln(\tau_{\rm coh} \Delta) /\lambda_M]/W$.

\textit{\blue{Vortex dynamics.}} When the width of the sample exceeds the superconducting coherence length, vortices in the $x-y$ plane of the superconducting film become well-defined topological excitations, and the supercurrent acquires an additional decay channel related to vortex tunneling across the sample. This effect therefore relies on the dynamics of vortices, described by the action:
\begin{equation}
S_{\rm kin} = \int d\tau \left(\frac{m \dot{\v x}^2}{2} + i\, \alpha\, \dot x\, y\right) +\eta \int \frac{d\omega}{2\pi} \vert \omega \vert  \vert \v x(\omega)\vert^2,
\end{equation}
where $\v x(\tau) = (x(\tau),y(\tau))$ is the time-dependent position of a given vortex.
The first term represents the kinetic energy for a point particle of mass $m$, the term proportional to $\alpha$ describes the Magnus force, which is somewhat similar to the Lorentz force in a magnetic field, and the last term, proportional to $\eta$, describes dissipation. 

The values of these parameters drastically depend on whether the vortex core is featureless, or whether it contains a quasi-continuum of bound states. In the first case, dissipation is absent ($\eta = 0$) and the parameter $\alpha$ is topologically quantized (it is given by the superfluid density).\cite{AoThouless1993} The mass diverges logarithmically if the superfluid is neutral, but the logarithm is cut inside the core for charged superconductors, where the mass is therefore minute, $m \sim m_{\rm electron} \lambda_{\rm TF}/\xi $ (here, $\lambda_{\rm TF}$ is the Thomas-Fermi screening length).\cite{Suhl1965,Brandt1977,DuanLeggett1992,Duan1993,DemircanNiu1996}
 
Here, we concentrate on the more realistic second case, in which the Caroli-deGennes-Matricon-type~\cite{CaroliMatricon1964} bound states inside the vortex core cannot be neglected. In $s$-wave superconductors, the spacing $\omega_0$ of these subgap states~\bibnote{In what follows, a possible ``zero-bias anomaly'' structure in the density of the core bound states (peak or dip, depending on the strength of disorder \cite{CaroliMatricon1964,Trivedi2021}) 
will be inessential for our analysis.} can be estimated to be $\omega_0 \sim \nu^{-1} \xi^{-2}$, where $\nu$ is the metallic density of states. Typically, 
in the 
clean (dirty) 
limit, 
$\omega_0 \sim\Delta^2/E_F$ 
($\omega_0 \sim\Delta/E_F \tau_{\rm el}$) 
is negligible
compared to $1/\tau_{\rm el}$. For a moving vortex that explores different microscopic realizations of disorder, this leads to a quasi-continuum of states, which allows for energy dissipation, yielding $\eta \sim n \omega_0 \tau_{\rm el}$, where $n$ is the density of electrons in the normal state.~\cite{BardeenStephen1965,FeigelmanLevit1993} At the same time, the Magnus force acquires a second topological contribution resulting from the spectral flow of bound states, which is equal in magnitude but opposite to the hydrodynamic contribution discussed above, so that $\alpha$ effectively vanishes.~\cite{KopninKravtsov1976,VolovikBook} Finally, the vortex mass is given by the total mass of particles trapped inside the vortex, $m \sim m_{\rm electron} \xi^2/\lambda_F^2$.~\cite{VolovikBook}

\begin{figure}
\includegraphics[width = .48 \textwidth]{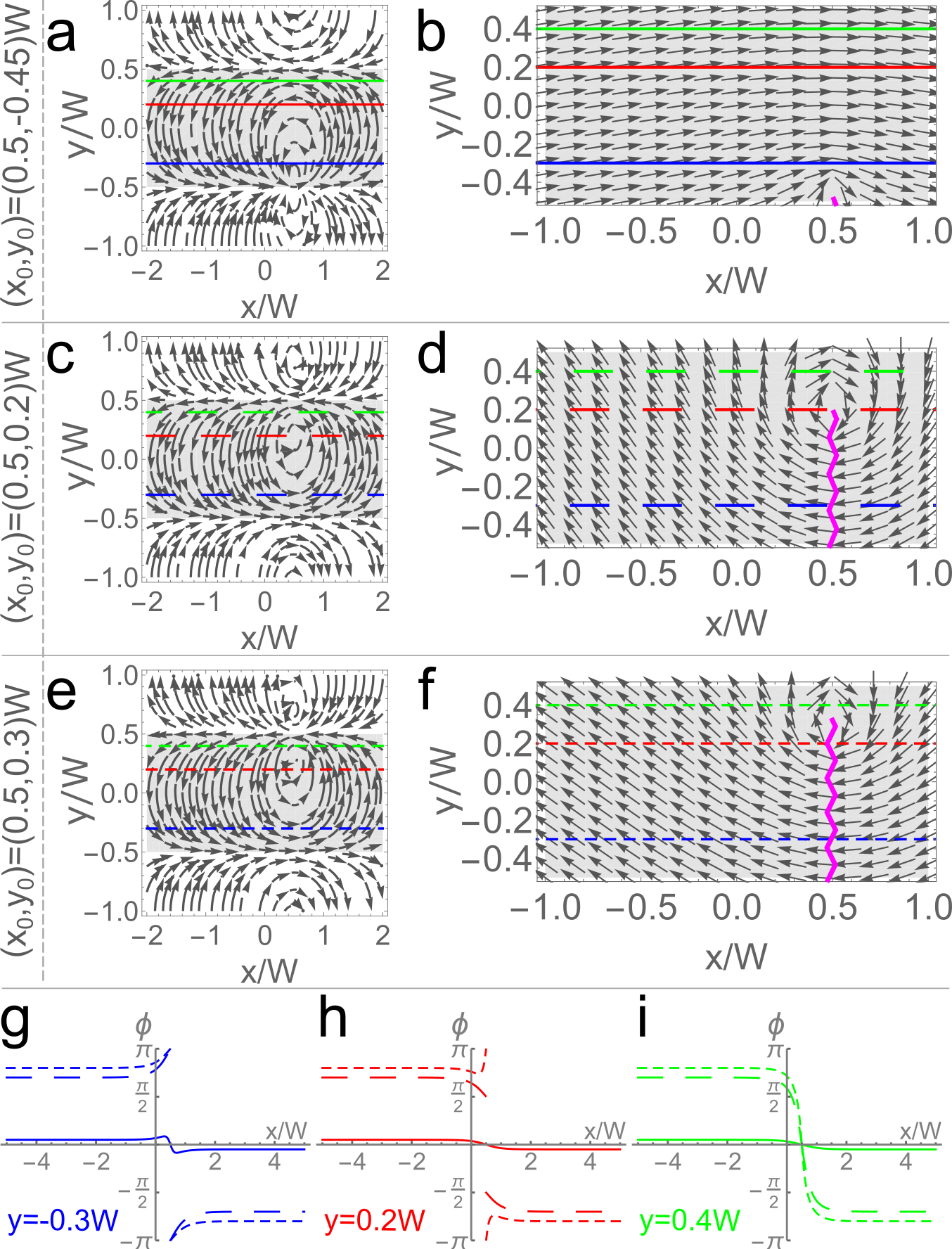} 
\caption{Cartoon of a single vortex at position $(x_0,y_0)$ tunneling from lower to the upper edge. (a),(c),(e): Gradient of the superconducting phase $\nabla \phi$. Note that mirror charges outside the sample (gray shaded) ensure no current outflux. (b),(d),(f): Vector field $(\cos (\phi), \sin(\phi))$. Our choice for the branch cut is represented by a pink zig-zag line. (g), (h), (i): Principal branch of the phase $\phi$ along the colored horizontal lines of panels (a)-(f) (same color and dashing code).}
\label{fig:UnbindingCartoon}
\end{figure}

\textit{\blue{Single vortex tunneling.}} 
Let us now inspect tunneling events involving a single vortex in the system. We remind the reader of the logarithmic attraction of a vortex-antivortex pair, which physically reflects the principle of minimizing the region of costly superflow between defects of opposite winding. It is, perhaps, less known that a single vortex in a finite superconducting strip is attracted to the boundaries. Technically, this results from the implementation of no-current-outflux boundary conditions by means of mirror charges (see Fig.~\ref{fig:UnbindingCartoon}a,c,e and Supplement~\cite{Supplement}
) and leads to a potential 
$V(y) = 2J\ln[W{\cos(\pi y/W)}/{\xi}]$, 
that should be included in the total action of the vortex in the strip geometry.
Here, $J$ is the 2D superconducting stiffness: $J \sim E_F$ ($J \sim E_F \Delta \tau_{\rm el}$) in the clean (dirty) limit at zero temperature. 

We begin by considering a `kink solution', $y_{\rm kink}(\tau)$, of the equation of motion, i.e., the unbinding of a single vortex from lower boundary and subsequent tunneling across the system, see Fig.~\ref{fig:UnbindingCartoon} for illustration. 
The tunneling action is $S_{\rm kink} (W) = S_{\rm pot}(W) +  S_{\rm cond}(W) $, where
\begin{align}
S_{\rm pot} &= \int_{-W/2+\xi}^{W/2-\xi}\! dy  \sqrt{2mV(y)} =  2 W \sqrt{m J} f(\pi \xi/W), \label{eq:aProcess}
\end{align}
is the contribution from the potential barrier. Here, we have introduced the dimensionless function $f(x)$ that satisfies
{$f(x)\simeq\sqrt{\ln(1/x)}$ in the limit $x\ll 1$ and $f(\pi/2)=0$. 
The second contribution, $S_{\rm cond}  = E_{\rm cond} \mathcal T$, is the condensation energy $E_{\rm cond}\sim J$ times the total length of the worldline $\mathcal T = \mathcal T(W)$. Here, $\mathcal T(W) = W/[v \sqrt{\ln(W/\pi\xi)}]$} for wide strips $W \gg \xi$ (with $v = \sqrt{J/m} \sim v_F \Delta/E_F$ the typical vortex speed), whereas for $W \rightarrow 2\xi$,  $\mathcal T(W) \rightarrow 1/\Delta$, which is the time for a vortex to nucleate. Thus, for narrow strips, ${S_{\rm kink}(W\rightarrow 2\xi)\simeq~J/\Delta} \sim K_{\rm 1D}\vert_{W \sim \xi}$ approaches the same value as the core action of a quantum phase slip, while for wide strips, both $S_{\rm pot}(W)$ and $S_{\rm cond}(W)$ are of the order of $J W/v \sim (W/ \lambda_F) E_F \min(\tau_{\rm el}, 1/\Delta)$, which is $(\xi/\lambda_F)$ times larger than the core action of a phase slip in Eq.~\eqref{eq:Bounce}.

A finite current bias lifts the degeneracy of the potential minima at $y = \pm W/2$ by an additional term $V_I = - \Phi_0 I y/W$. As mentioned, the supercurrent decay is governed by a bounce solution,~\cite{Coleman1977,Kleinert2009} which we approximate by a kink and an antikink at temporal distance $\delta \tau$: $y_{\rm bounce}(\tau) = y_{\rm kink}(\tau + \delta \tau/2) - y_{\rm kink}(\tau - \delta \tau/2) + W/2 -\xi$. Using this Ansatz, we obtain the following $\delta \tau$-dependent action, see Supplement~\cite{Supplement}, 
\begin{align} \label{eq:Bounce2D}
S_I (\delta \tau) &= 2 S_{\rm kink}(W) + S_{\rm int}(\delta \tau) - \Phi_0 I \delta \tau.
\end{align}

At variance with the 1D bounce solution, Eq.~\eqref{eq:Bounce}, the attraction between kink and antikink, encoded in $S_{\rm int}(\delta \tau)$, can not be determined exactly. It asymptotically vanishes if the vortex disappears at the opposite boundary,\bibnote{Here, we do not describe microscopically the process of vortex disappearance at the boundary, assuming that the relaxation described by $\tau_\text{coh}$ is sufficient for transferring the core electrons to the condensate or to the leads.} 
$S_{\rm int}(\delta \tau) \to 0$ for $\delta \tau > \mathcal T$ and monotonically increases in the preceding regime $0< \mathcal T - \delta \tau  \ll \mathcal T$, where $S_{\rm int}(\delta \tau) \simeq -E_{\rm cond}(\mathcal T - \delta \tau)$, which shows that the dominant reason for attraction is the string tension of the vortex worldline.  

Formally, the instanton contribution to the partition sum can be evaluated using the steepest-descent method analogously to the bounce action \eqref{eq:Bounce}. However, the properties of $S_{\rm int}(\delta \tau)$ imply that the typical dwell time $\delta \tau$ diverges as a function of $I \rightarrow 0$, exceeding the coherence time $\tau_{\rm coh} \gg \mathcal T$. Consequently, evaluation of Eq.~\eqref{eq:VoltageDecayrate} similarly to \eqref{eq:RSlips} but using Eq.~\eqref{eq:Bounce2D} instead of Eq.~\eqref{eq:Bounce} yields
\begin{equation}\label{eq:RVortex}
R_{\rm {SVT}} = \frac{h}{e^2} \frac{L}{\xi}
\mathcal{A}_{\rm vortex}\ e^{- 2 S_{\rm kink}(W)}.
\end{equation}
The pre-exponential factor $\mathcal{A}_{\rm vortex}$ depends on the nature of relaxation and should be evaluated for a given microscopic model of relaxation. 

We conclude the discussion of a single-vortex tunneling with three comments. First, the logarithmic vortex interaction is cut beyond the Pearl length\cite{Supplement} $\lambda_M$
so that Eq.~\eqref{eq:aProcess} is valid only for samples $W < \lambda_M$. Second, the effect of dissipation on the tunneling has been disregarded here, which is valid if the tunneling time $\mathcal T$ is small compared to the dissipation time $m/\eta \sim 1/\omega_0^2 \tau_{\rm el}$. This is equivalent to the condition $W \ll \text{min}(v_F \tau_{\rm el}, \xi)/(\omega_0 \tau_{\rm el})$. We will see in the next section that the resistance is dominated by vortices tunneling across an effective distance $d_{\rm opt} \ll W$, which satisfies both the bounds imposed by screening and by dissipation, even for wide junctions. Third, in deriving the effective tunneling action, we disregarded the mesoscopic fluctuations of the superconducting gap (and, hence, stiffness $J$). \cite{SkvortsovFeigelman2005,KoenigMirlin2015}
As we demonstrate in Ref.~[\onlinecite{Supplement}] 
for $E_F \tau_{\rm el}\gg 1$, these fluctuations do not affect the exponential factor in Eq.~(\ref{eq:RVortex}).

\begin{figure}
\includegraphics[width = .48 \textwidth]{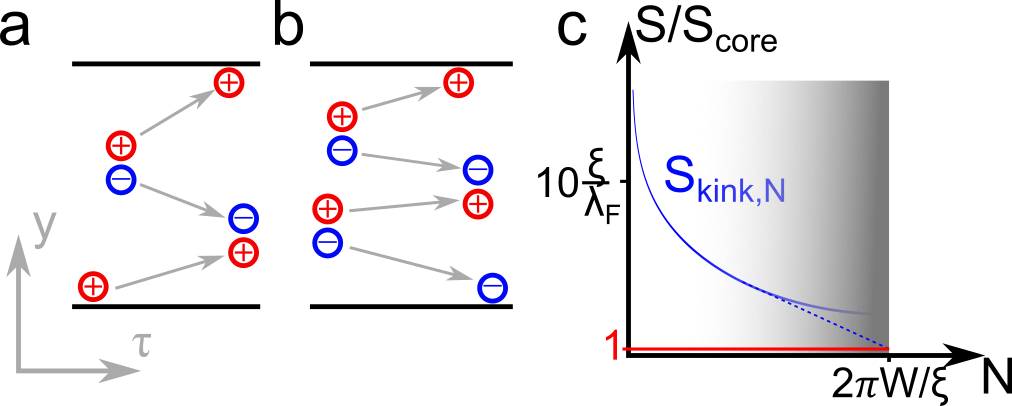} 
\caption{Multi-vortex tunneling processes. (a) The unbinding of a vortex from the lower edge, similar to Fig.~\ref{fig:UnbindingCartoon}, but assisted by an additional dipole nucleating in the bulk of the system. (b) The decay of two vortex dipoles into a dipole and two vortices disappearing at the edge. (c) The tunneling action (in units of $S_{\rm core}$ introduced in Eq.~\eqref{eq:Bounce}) for a multi-vortex tunneling process as a function of $N$.}
\label{fig:MultiVortex}
\end{figure}

\textit{\blue{Multi-vortex tunneling.}}  
A simple tunneling event that involves more than one vortex is a dipole dissociation,\cite{IengoJug1995} in which a quantum fluctuation creates a dipole of vortices inside the strip and subsequently the dipole constituents tunnel towards opposite edges, leading to an overall phase slip of $2\pi$. 
Here, we consider a generalization of this event where $n$ dipoles are nucleated in the strip and subsequently tunnel to either side of the strip, see Fig.~\ref{fig:MultiVortex}b. We also consider a generalization of the edge unbinding event discussed above, Fig.~\ref{fig:UnbindingCartoon}, where the tunneling of a vortex from one edge to the other is assisted by $n$ dipoles in the bulk, see Fig.~\ref{fig:MultiVortex}a. 

As a natural Ansatz, we consider (see Supplement\cite{Supplement}
) the total action $S_{\rm kink,N} =\sum_{i = 1}^N S_{\rm kink}(d_i)$, where $S_{\rm kink}(d)$ is the kink action for a single vortex tunneling in a strip of size $d$ introduced before Eq.~\eqref{eq:aProcess}, $d_i$ is the tunneling distance of the $i$th vortex, and $N = 2n +1$ in the case of Fig.~\ref{fig:MultiVortex}a and $N = 2n$ in case of Fig.~\ref{fig:MultiVortex}b.
The minimization of the action prescribes that all dipoles nucleate at the same moment in time and same $x$-position, and that the nucleation points are furthermore equally spaced in $y$-direction, i.e., each vortex travels the same distance $d_N= W/N$, so that the kink action is $S_{\rm kink,N} = N S_{\rm kink}(d_N)$. We remark that this is an accurate tunneling action for the optimum event; yet, the Ansatz that we used is heuristic and employing it for the calculation of fluctuation determinant is not exact, although parametrically correct.

Increasing $N$ in $S_{\rm kink,N}$ has two effects. On one hand, $N S_{\rm pot}(d_N)$ decreases with the number of dipoles. On the other hand, the contribution of the string tension $N E_{\rm cond} \mathcal T (d_N)$ increases with $N$ (even though $\mathcal T (d_N)$, of course, decreases). The first effect prevails, see Fig.~\ref{fig:MultiVortex}c, in which the solid curve is obtained on the basis of the vortex motion at length scales large compared to $\xi$, and the dashed curve is the extrapolation of $\mathcal T (d_N) \searrow 1/\Delta$ as $d_N$ approaches $\xi$ (shaded region).  
The action is monotonically decreasing with increasing number of vortices, leading to an optimum number $N_{\rm opt} \sim W/\xi$ and $d_N \sim 2\xi$.  The resistance produced by multi-vortex configurations,
\begin{equation}\label{eq:MultiVortexResistance}
R_{\rm MVT} = \frac{h}{e^2} \frac{L}{\xi} \mathcal{A}\ e^{-2 (W/\xi) S_{\rm kink}(2\xi)}
\end{equation}
follows along the lines of the calculation for the single-vortex tunneling event (the preexponential factor $\mathcal{A}$, which may depend on relaxation details, is considered separately). 
Clearly, as $d_N\sim\xi$, vortices cease being well-defined excitations, and the multi-vortex tunneling event cannot be distinguished from a quantum phase slip. It is, therefore, a reassuring consistency check that the exponent in the resistance due to $N$ co-tunneling vortices, Eq.~\eqref{eq:MultiVortexResistance}, $N S_{\rm kink}(2\xi)\sim NJ/\Delta\sim W E_F \min(1/\Delta, \tau_{\rm el})/\xi \sim K_{\rm 1D}$, is in accordance with the exponent of the resistance due to quantum phase slips, Eq.~\eqref{eq:RSlips}.

\textcolor{blue}{\textit{Conclusion.}}  In summary, we have investigated the resistance of superconducting strips with moderate disorder $1/\tau_{\rm el} \gg \Delta^2/E_F$ near zero temperature.
The obtained resistance $R$ is dominated by cotunneling of $N \sim W/\xi > 1$ vortices, a process that is equivalent to a quasi-1D quantum phase slip.
This yields $\ln R \sim - (W/\xi) E_F \min (1/\Delta, \tau_{\rm el})$, see Eqs.~\eqref{eq:RSlips} and \eqref{eq:MultiVortexResistance}, which 
qualitatively agrees with the low-$T$ saturation values of resistance presented in Fig.~1 of Ref.~[\onlinecite{SchneiderStrunk2019}].
Our result for exponential dependence of resistance on $W$ is reminiscent of exponential suppression of conductance with the system size on the insulating side of SIT.

While we concentrated on the exponent of the superconducting decay rate, we conclude with a  discussion of the preexponential factor. {Its $W$-dependence stems from quantum fluctuations of the tunneling trajectory, only, if the mechanism of dissipation is local.} 
Using the above-mentioned Ansatz for multi-vortex events, we estimate that each of the $W/\xi$ vortices contributes to $\mathcal{A}$ in Eq.~\eqref{eq:MultiVortexResistance} a fluctuation determinant of the order of the inverse kink action
,\cite{Supplement} leading to a preexponential factor 
$[S_{\rm kink}(2\xi)]^{-W/\xi}$. (A similar factor is expected to arise from spatial fluctuations about the straight $y$-independent phase-slip world-line.) By re-exponentiation, this quantum-fluctuation-induced prefactor can be viewed as an entropic contribution to the tunneling action, leading to
\begin{equation}\label{eq:LogRFinal}
\ln(e^2R/h) \sim -W[g + \ln(g)]/\xi + \ln(L/\xi) \,,
\end{equation} 
where, in the dirty limit, $S_{\rm kink}(2\xi) \sim E_F \tau_{\rm el} \equiv g$ is given by the normal-state conductance, see Fig.~\ref{fig:Setup}c.
Pushing our theory to the limit of its applicability, 	
we observe a sign change of the action at $g\sim 1$, reminiscent of SIT.

We acknowledge useful discussions with D.~Bagrets, A. ~Bezryadin, I.~Burmistrov, A.~Finkelstein, E.~Khalaf, A.~Rogachev, D.~Shahar, M.~Skvortsov, C.~Strunk, H.~Xie, and A.~Zaikin. Support for this work at the University of Wisconsin-Madison (A.L.) was provided by the National Science Foundation, 
Quantum Leap Challenge Institute for Hybrid Quantum Architectures and Networks, NSF Grant No. 2016136.


\bibliography{FiniteSizeSCBIB}

\begin{thebibliography}{66}
\bibitem [{3()}]{3}%
  \BibitemOpen
  \href@noop {} {}\bibinfo {note} {Here, we do not describe microscopically the
  process of vortex disappearance at the boundary, assuming that the relaxation
  described by $\tau _\text {coh}$ is sufficient for transferring the core
  electrons to the condensate or to the leads.}\BibitemShut {Stop}%
\end{thebibliography}

\clearpage

\setcounter{equation}{0}
\setcounter{figure}{0}
\setcounter{table}{0}

\makeatletter
\renewcommand{\theequation}{S\arabic{equation}}
\renewcommand{\thefigure}{S\arabic{figure}}
\renewcommand{\thesection}{S-\Roman{section}}
\renewcommand{\bibnumfmt}[1]{[S#1]}
\begin{widetext}
\setcounter{page}{1}
\begin{center}
Supplementary materials on \\
\textbf{"Resistance of 2D superconducting films"}\\
Elio J. K\"onig$^{1}$, I.~V.\ Protopopov$^{2,3}$, A. Levchenko$^{4}$, I.V. Gornyi$^{5,6,7}$, {A.~D.\ Mirlin}$^{5,6,3,8}$\\ 

$^{1}${Max Planck Institute for Solid State Research, D-70569 Stuttgart, Germany}\\
$^{2}${Department of Theoretical Physics, University of Geneva, 1211 Geneva, Switzerland}\\
$^{3}${L.~D.\ Landau Institute for Theoretical Physics RAS, 119334 Moscow, Russia}\\
$^{4}${Department of Physics, University of Wisconsin-Madison, Madison, Wisconsin 53706, USA}\\
$^{5}${Institute for Quantum Materials and Technologies, Karlsruhe Institute of Technology, 76021 Karlsruhe, Germany}\\
$^{6}${Institut f\"ur Theorie der kondensierten Materie, Karlsruhe Institute of Technology, 76128 Karlsruhe, Germany}\\
$^{7}${Ioffe Institute, 194021 
 St.~Petersburg, Russia}\\
$^{8}${Petersburg Nuclear Physics Institute, 188300 St.~Petersburg, Russia} 
\end{center}
\end{widetext}

These supplementary materials contain an overview of list of scales, Sec.~\ref{app:Scales}, employed in the paper, a derivation of the vortex interaction in the strip with and without screening (Secs.~\ref{sec:model}, \ref{app:Screening}, respectively), details about vortex tunneling, Sec.~\ref{app:VortexTunneling}, about the multivortex tunneling configurations, Sec.~\ref{app:MultivortexAction} and about the effect of mesoscopic fluctuations, Sec.~\ref{App:Disorder}.

\section{List of scales}
\label{app:Scales}

\subsection{List of energy and time scales}

\begin{itemize}
\item elastic scattering time $\tau_{\rm el}$.
\item coherence time $ 1/\Delta$
\item Tunneling time $\mathcal T$ 
\item 2+1 D stiffness $J$ at zero temperature. Clean case: $J \sim E_F$. Dirty case: $J = \Delta E_F \tau_{\rm el}$.  
\end{itemize}

\subsection{List of length scales}

\begin{itemize}
\item Fermi wave length $\lambda_F$
\item UV length scale $\lambda_0$. Clean case: $\lambda_0 =\lambda_F$. Dirty case: $\lambda_0 = \lambda_F/\sqrt{\Delta \tau_{\rm el}}$
\item Thomas Fermi wave length $\lambda_{\rm TF}$
\item Coherence length $\xi$. Clean case: $\xi = v_F/\Delta$, dirty case: $\xi = v_F \sqrt{\Delta/\tau_{\rm el}}$.
\item Pearl length $\lambda_M= \lambda_{\rm TF} c^2/v_0^2$.
\end{itemize}

\section{Vortex interaction parameters and screening}
\label{sec:model}

For the convenience of the reader, this section summarizes basic properties of vortex interactions in the finite strip.

\subsection{Long-wavelength action}

We consider the following effective long-wavelength action of the superconducting plate, see Fig.~\ref{fig:Setup}
\begin{equation}
S = \int_{(x, y, \tau)} \left [ \frac{J}{\pi} \left (\frac{2e}{c} \v A - \nabla \phi \right )^2 + \frac{Z}{\pi} \left (2 e A_0 + \partial_\tau \phi\right )^2 \right ], \label{eq:O2action}
\end{equation}
where $\int_{(x, y, \tau)}  = \int_{-L/2}^{L/2} dx \int_{-W/2}^{W/2} dy \int_{-\infty}^{\infty} d\tau$, supplemented with the Maxwell action of electromagnetic fields $(A_0,\vec A)$ living in (3+1) dimensional space-time. 
As the phase field $\phi$ may contain singularities it is important to keep in mind that $\vec{\nabla} \phi$ is in general \textit{not} a gradient field. The notation used in Eq.~\eqref{eq:O2action} is suggestive and mathematically correct only away from the singularities.

The parameters $J,Z$ of this action define a characteristic length scale $\lambda_0 = 1/\sqrt{J Z}$ and speed $v_0 = \sqrt{J/Z}$.\cite{AbrahamsTsuneto1966,Stoof1993} In addition, they define length scales associated to the screening of electric charge $\lambda_{\rm TF} = 1/[16 {e^2} Z] \sim \lambda_0 v_0/[\alpha_{\rm QED} c]$ (``Thomas-Fermi length'') and current $\lambda_M = c^2/[16 e^2 J] = \lambda_{\rm TF} {c^2}/{v_0^2} \sim \lambda_0 c/[\alpha_{\rm QED} v_0]$ (``Pearl-London length''), where $1/\alpha_{\rm QED} \approx 137$ if the sample is suspended in vacuum. 

In our calculations, we use $v_0 \ll c$ and additionally assume the following realistic hierarchy of length scales $ L \gg \lbrace W, \lambda_M \rbrace \gg \xi \gg \lbrace \lambda_{\rm TF}, \lambda_0 \rbrace $. Note that, in typical experimental samples, the Pearl-London length, which is about four to five orders of magnitude larger than $\lambda_0 \gtrsim 1 nm$, can be smaller, larger or comparable to the system's width.

Since we are considering a charged superconductor, the smooth part of the Goldstone mode $\phi$ can be reabsorbed in a redefinition of the electromagnetic fields $(A_0,\v A)$ (``Anderson-Higgs-mechanism''). However, singular configurations of $\phi$, such as vortices, and an externally applied current bias can not be ``eaten'' by a gauge transformation of the vector potential. Differentiation of the action \eqref{eq:O2action} with respect to the gauge potentials $(A_0,\v A)$ implies the 2D charge and current densities, $\rho$ and $\v j_{\rm tot} = \v j + j_0 \hat e_x$, respectively, where 
\begin{subequations}
\begin{eqnarray}
\rho &=& - 4 e \frac{Z}{\pi} \left (\partial_\tau \psi + 2 e A_0\right ), \\
\v j &=& \frac{4 e}{c} \frac{J}{\pi} \left (\nabla \psi - \frac{2 e}{c} \v A\right ), \label{eq:currentdensity} 
\end{eqnarray}
and 
\begin{equation}
 j_0 \equiv \frac{I}{W c} = const.
\end{equation}
\end{subequations}
is the externally injected, fixed background current density. Here, we have separated the phase into two contributions
\begin{equation}
\phi =  \psi +\phi_I
\end{equation}
where the time independent field $\phi_I$ enters $j_0$ and $\psi$ represents the fluctuating part of the phase field. The finite geometry implies the boundary conditions
\begin{equation}
\left .j_x \right  \vert_{x= \pm L/2} = 0, \;\left .j_y \right  \vert_{y= \pm W/2} = 0. \label{eq:boundarycond}
\end{equation}
We will be interested in the limit of linear response in $I$.

\subsection{Vortex configurations.}
\label{sec:VoltageGeneration}

As mentioned in the introduction, the finite voltage in superconductors relies on quantum phase slips and tunneling of vortex excitations. Here, we provide further details leading to the technical definition of the problem.
Generally, in a 2D system characterized by coordinates $(x_1,x_2)$, a vortex of winding $n_w$ at $(x_{1,0},x_{2,0})$ is defined by the singular field configuration
\begin{align}
\psi_{n_w,x_{1,0},x_{2,0}}(x_1,x_2) &= n_w \arctan\left [\frac{x_2 - x_{2,0}}{x_1 - x_{1,0}}\right ] \notag \\
&+ \pi \theta(x_{1,0} - x_1).\label{eq:generalVortexDef}
\end{align}
This mapping represents the principal branch of the phase field. In our convention, the branch cut is chosen on the line $x_1 = x_{1,0}$, $x_2 < x_{2,0}$. In the limit $\lambda_M \gg W$ vortex field lines in a charged superconductor are the same as in a neutral superfluid. A vortex field complying with the no out flux boundary conditions~\eqref{eq:boundarycond}, can be constructed by an infinite series of mirror vortices and has the form,\cite{OvchinnikovVarlamov2015} (see Sec.~\ref{app:VortexPotScreening})
\begin{eqnarray}
\psi^{\rm (M)}_{n_w,x_0,y_0}(x,y) &=& n_w \Big \lbrace \arctan\left [\frac{\tan\left (\frac{\pi}{2W} (y -y_0)\right )}{\tanh \left (\frac{\pi}{2W}(x-x_0)\right )}\right ] \notag \\
&&-   \arctan\left [\frac{\tan\left (\frac{\pi}{2W} (y + y_0-  W)\right )}{\tanh \left (\frac{\pi}{2W}(x-x_0\right )}\right ]\notag \\
&&- \pi \theta(y+y_0) \text{sign}(x-x_0)\Big \rbrace. \label{eq:MirrorVortex}
\end{eqnarray}
This configuration is shown in Fig.~\ref{fig:UnbindingCartoon}.

\subsection{Vortex interactions}
\label{app:VortexIA}
The definining property
\begin{equation}
\nabla \times (\nabla \psi^{\rm (M)}_{n_w,x_0,y_0}(x,y)) = 2\pi n_w \delta(\v x - \v x_0),
\end{equation}
readily leads to the single vortex potential presented in the main text
, i.e. a logarithmic attraction
\begin{equation}
V(\v x) \simeq 2 J \ln(\vert \v x \vert/\xi)
\end{equation}
for a vortex -antivortex pair which is at distance $\v x$.

The trigonometric function in the potential of a single vortex in a finite strip 
can be derived from Eq.~\eqref{eq:MirrorVortex} or as follows: Consider a sequence of vortices and anti-vortices in an infinite plane at positions
\begin{equation}
y_+^{(i)} = i d - y, \quad
y_-^{(i)} = (i-1) d + y, \quad y \in (0,d/2), i \in \mathbb Z.
\end{equation}
For $d = W/2$ and $y= y_0 + d/4$ this sequence corresponds to the sequence of mirror charges of a single vortex in a strip of width $W$ and at position $y_0$.
The total potential energy is 
\begin{equation}
V = 2J \sum_{i >j} \ln\left[\frac{y_+^{(i)} - y_-^{(j + 1)}}{y_-^{(i)}-y_-^{(j)}} \frac{y_-^{(i)} - y_+^{(j)}}{y_+^{(i)}-y_+^{(j)}}\right].
\end{equation}
This obviously diverges linearly with the number of dipoles. The energy per dipole is
\begin{eqnarray} \label{eq:VperDipole}
V_{\text{per dipole}} &=& 2J \sum_{n > 1} \ln \left [\frac{(n - 1/2)^2 + (2 y/d -1/2)^2}{n^2} \right] \notag \\
&=& 2 J \ln \left [\sin\left (\frac{2\pi y}{d} \right) \right] + J \times \text{const.}.
\end{eqnarray}
The constant term is $y$ and $d$ independent and thus physically irrelevant, we therefore do not discuss possible regularization schemes of this logarithmically divergent term. Clearly, this reproduces the potential for a single vortex in a finite strip as given in the main text, 
\begin{align} \label{eq:SingleVortexPot}
V(y_0) &= 2J  \ln \left [\frac{\cos(\pi y_0/W)}{\xi/W}\right ].
\end{align}
We also use Eq.~\eqref{eq:SingleVortexPot} for the calculation of multivortex tunneling actions.

\section{Screening of vortices via London electromagnetism}
\label{app:Screening}

We here briefly summarize the effect of screening of vortices and phase slips and determine that none of these effects is important in the parameter regime of interest. 
We largely follow \cite{DuanLeggett1992, Duan1993} and omit retardation effects in view of the small parameter $v_0/c$.

Since the action of electromagnetic fields is Gaussian, integration of electromagnetic fields is equivalent to the saddle point treatment dictated by the Maxwell equations (we use Lorenz gauge),
\begin{subequations}
\begin{eqnarray}
\lbrace - \partial_{c \tau}^2 - \nabla_{3D}^2 \rbrace A_0 &=& 4 \pi \rho  W\tilde \delta(y) \delta(z), \\
\lbrace - \partial_{c \tau}^2 - \nabla_{3D}^2 \rbrace \v A &=& 4 \pi \v j W \tilde \delta(y) \delta(z). \label{eq:MaxwellEquations:Elfield}
\end{eqnarray}
\label{eq:MaxwellEquations}
\end{subequations}
We introduced the notation $\tilde \delta(y) = \theta(y+W/2)\theta(W/2-y)/W$, where $\theta(x)$ is the Heaviside function. Since the background current $j_0$ is kept fixed, it should not be further screened and does not enter these Maxwell's equations. Integration of electromagnetic fields leads to the effective action $S= S_0 + S_I$ of the matter field\cite{DuanLeggett1992,Duan1993}
\begin{subequations}
\begin{eqnarray}
S_0 &=& \int_{(x,y,\tau) }\Big[ \frac{J}{\pi} \nabla \psi \left (\nabla \psi-\frac{2e}{c} \v A  \right ) \notag \\
&& + \frac{Z}{\pi}\partial_\tau \psi \left (\partial_\tau \psi+2 e A_0 \right ) \Big ], \label{eq:O2action:afterintegration} \\
S_I &=& 2  \int_{(x,y,\tau) } \frac{c j_0}{2e} \left (\partial_x \psi - \frac{2e}{c} A_x\right ).\label{eq:O2action:SI}
\end{eqnarray}
\end{subequations}
The electromagnetic fields entering Eq.~\eqref{eq:O2action:afterintegration} are solutions to Eq.~\eqref{eq:MaxwellEquations} for a given field configuration $\psi$. The regular part of $\psi$ can be removed by partial integration combined with the continuity equation (gauge invariance).

\subsection{Screening of the phase-slip}
\label{sec:ScreeningPhaseslip}

We consider a phase slip in the phase field $\psi$ at space-time $ (x_0, \tau_0)$.
The gradient field is ($\tilde \tau = v_0 \tau$)
\begin{equation}
\nabla \psi(\underline x)= \left (\begin{array}{c}
\partial_x \\ 
\partial_{\tilde \tau}
\end{array} \right ) \psi = \frac{1}{x^2 + \tilde \tau^2}\left (\begin{array}{c}
- \tilde \tau \\ 
x
\end{array} \right ) 
\end{equation}
with $\underline x^T = (x, \tilde \tau)$.
In Fourier space we find
\begin{equation}
(\nabla \psi) (\v q) = - \frac{2\pi i}{\v q ^2} \left (\begin{array}{c}
- q_{\tilde \tau} \\ 
q_x
\end{array} \right )
\end{equation}
where $\v q = (q_x, q_{\tilde \tau})$.

We 4D Fourier transform Eq.~\eqref{eq:MaxwellEquations} using $\tilde \delta(y) = \delta(y)$ and introduce $\tilde{\underline{A}}_{1D} (\v q)=  \underline A (q_x, y = 0, z = 0, q_{\tilde \tau})$, use the integral
\begin{equation}
\int \frac{dq_y dq_z}{(2\pi)^2} \frac{1}{q_x^2 + q_y^2 + q_z^2} = \frac{ \ln \vert \frac{1}{q_x r_x}\vert}{2\pi} 
\end{equation}
with $r_x \sim \xi$ the extension of the asymmetric vortex core in $x-$direction to obtain

\begin{subequations}
\begin{eqnarray}
\left (  \frac{2\pi}{ \ln \vert \frac{1}{q_x r_x}\vert} + 2 \pi \gamma_E  \right ) \tilde A_{0,1D} &=&- \frac{\pi \gamma_E v_0}{e} \partial_{\tilde \tau} \psi (\v q) \\
\left ( \frac{2\pi}{ \ln \vert \frac{1}{q_x r_x}\vert} + 2 \pi \gamma_M  \right ) \tilde A_{x,1D} &=& \frac{\pi \gamma_M c}{e} \partial_x \psi  (\v q).
\end{eqnarray}
\label{eq:London}
\end{subequations}
Here, $\gamma_{M,E} = W/\lambda_{M,E}$. This leads to
\begin{eqnarray}
 [\partial_x \psi - \frac{2e}{c} A_x](\v q) &=& \frac{\gamma_M^{-1}}{\gamma_M^{-1} + \ln \vert \frac{1}{q_x r_x} \vert} \partial_x \psi(\v q)\\
 \frac{1}{v_0} [\partial_\tau \psi + {2e} A_0](\v q) &=& \frac{\gamma_E^{-1}}{\gamma_E^{-1} + \ln \vert \frac{1}{q_x r_x} \vert} \partial_{\tilde{\tau}} \psi(\v q).
\end{eqnarray}
We observe that screening is important for $ -\ln \vert{q_x r_x} \vert > \gamma_{M,E}^{-1}$ and thus only for large systems $L > \xi e^{\gamma_M^{-1}}$.

Anticipating, that the saddle point configuration of the dipole occurs at $x_+ = x_-$ we obtain
\begin{eqnarray}
S_{\rm IA} &=& \frac{2 \tilde J W}{\pi} \int dq_x dq_{\tilde \tau} \Big \lbrace \frac{q_{\tilde \tau}^2}{\v q^4} \frac{\gamma_M^{-1}}{\gamma_M^{-1} + \ln \vert \frac{1}{q_x r_x} \vert} \notag \\
&& + \frac{q_{x}^2}{\v q^4} \frac{\gamma_E^{-1}}{\gamma_E^{-1} + \ln \vert \frac{1}{q_x r_x} \vert} \Big\rbrace \left [1- \cos(q_{\tilde \tau} \Delta \tilde \tau)\right ] \notag \\
&=&  \int_0^1 du \Big \lbrace \frac{1}{u}\frac{2 \tilde J W \gamma_M^{-1}}{\gamma_M^{-1} - \ln (u)} \left [1- e^{-\frac{\vert \Delta \tilde\tau \vert}{r_x} u} \left (1-\frac{\vert \Delta \tilde\tau \vert}{r_x} u \right ) \right ] \notag\\
&&+  \frac{1}{u}\frac{2 \tilde J W \gamma_E^{-1}}{\gamma_E^{-1} - \ln (u)} \left [1- e^{-\frac{\vert \Delta \tilde\tau \vert}{r_x} u} \left (1+\frac{\vert \Delta \tilde\tau \vert}{r_x} u \right ) \right ] \notag \\
&\simeq& 2 \tilde J W\Big \lbrace \gamma_M^{-1} \ln \left [1+ \gamma_M \ln\left (\frac{\vert \Delta \tilde\tau \vert}{r_x}\right )\right ] \notag \\
&&+ \gamma_E^{-1} \ln \left [1+ \gamma_E \ln\left (\frac{\vert \Delta \tilde\tau \vert}{r_x}\right )\right ] \Big \rbrace.
\end{eqnarray}
The symbol $\simeq$ means equality in the limit ${\vert \Delta \tilde\tau \vert}/{r_x} \rightarrow \infty$. We repeat that magnetic screening should only be kept provided $\gamma_M^{-1}= \lambda_M/W < \ln[L/\xi]$. 
So long as $\gamma_M \ln(\delta \tau v_0/\xi) \ll 1$, i.e. $\delta \tau \Delta \ll e^{\lambda_M/W}$ can be omitted, while electric screening effects can always be omitted. This concludes the derivation underlying our statement of the main text: Magnetic screening effects of quantum phase slips are negligible in the parameter regime of $W/\xi$, and lead to the double logarithm quoted in the main text.

\subsection{Screening of a vortex}
\label{app:VortexPotScreening}

We first discuss the screening of a static vortex configuration in the bulk of the system Eq.~\eqref{eq:MaxwellEquations} become
\begin{equation}
[- \nabla^2_{3D} + 2 \lambda_M^{-1} \delta(z)] \v A(\v x) = \frac{c \lambda_M^{-1}}{e} \nabla \psi(\v x) \delta(z).
\end{equation}
Fourier transformation  and the definition of $\v A_{2D} (\v q) = \v A (\v q,0)$ leads to
\begin{equation}
2 [\vert \v q \vert + \lambda_M^{-1}] \v A_{2D}(\v q) = \frac{c \lambda_M^{-1}}{e} \nabla \psi(\v q)
\end{equation}
and thus to
\begin{equation}
[\nabla \psi - \frac{2e}{c} \v A](\v q) = \frac{\vert \v q\vert }{\vert \v q\vert + \lambda_M^{-1}} \nabla \psi (\v q).
\end{equation}
In the case of a vortex, $\nabla_\mu \psi (\v q) = \epsilon_{\mu \nu} 2 \pi i q_\nu/\v q^2$, we can write after inverse Fourier transformation
\begin{subequations}
\begin{equation}
[\partial_\mu \psi - \frac{2e}{c} A_\mu](\v x) = - \epsilon_{\mu \nu} \partial_\nu F(r/\lambda_M)
\end{equation}
with 
\begin{eqnarray}
F(a) &=& - \frac{\pi}{2} \left (\mathbf H_0(a) - Y_0(a)\right ) \notag \\
 &\simeq & - \ln(1+ 1/a) + \frac{\gamma_{\rm EM}- \ln(2)}{1 + a^2}.
\end{eqnarray}
\end{subequations}
Here, $\gamma_{\rm EM}$ is the Euler-Mascheroni constant, $\mathbf H_0$ the Struve function of order zero, $Y_0$ the zeroth Bessel function of the second kind and the symbol $\simeq$ means asymptotic equality for both $a \gg 1$ and $a \ll 1$. Since only the derivative of $F$ enters into the action, we omit the second term, which has vanishing derivative in both limits.

It is also possible to calculate the resummation of image charges of a vortex at $(0,y_0)$ such that $\v j$ satisfies the boundary conditions. We define $F^{(M)}(x,y)$ via 
\begin{widetext}
\begin{eqnarray}
[\partial_y \psi - \frac{2e}{c} A_y] &=& \partial_x F^{(M)}(x,y) \notag \\
&=& \sum_k \Big [\frac{x}{x^2+(y-y_0 + 2 kW)^2} \frac{\lambda_M}{\lambda_M + \sqrt{ x^2+(y-y_0 + 2 kW)^2}} \notag \\
&& - \frac{x}{x^2+(y+y_0 + W - 2 kW)^2} \frac{\lambda_M}{\lambda_M + \sqrt{x^2+(y+y_0 + W - 2 kW)^2} } \Big ] \notag \\
&=& \int_{-\infty}^\infty du g(\sqrt{u^2 + \bar x^2}) \frac{\bar x}{\sqrt{u^2 +\bar x^2}} \left [ \delta(u) - \frac{\bar \lambda/\pi}{\bar \lambda^2 + u^2} \right ] \notag \\
&=& - \frac{2}{\pi} \int_0^\infty du \partial_u \left (\frac{\bar x}{\sqrt{u^2 +\bar x^2}} g(\sqrt{u^2 +\bar x^2}) \right ) \text{arccot} (u/\bar \lambda).
\end{eqnarray}
\end{widetext}
Here, the summation has been evaluated by a contour integral. We introduced the notation $\bar x = x/\pi W, \bar y = y/\pi W, \bar \lambda = \lambda_M/\pi W$ and 
\begin{eqnarray}
g(\bar x) &=& \frac{\sinh(\bar x) \pi/2W }{\cosh(\bar x) - \cos(\bar y - \bar y_0)} \notag \\
&&-  \frac{\sinh(\bar x) \pi/2W }{\cosh(\bar x) + \cos(\bar y + \bar y_0)} .
\end{eqnarray}
Note that $g(\bar x) \vert_{y_0 = \pm W/2} = 0 $, and thus $j_y \vert_{y_0 = \pm W/2} = 0$, in accordance with the boundary conditions.
We define 
\begin{eqnarray}
G(\bar x) &=& \frac{1}{2}  \Big ( \ln \left [\cosh(\bar x) - \cos(\bar y - \bar y_0) \right ] \notag \\
&& - \ln \left [\cosh(\bar x) + \cos(\bar y + \bar y_0) \right ] \Big )
\end{eqnarray} 
to obtain
\begin{equation}
F^{(M)} = - \frac{2}{\pi} \int_0^\infty du \partial_u G(\sqrt{u^2 + \bar x^2}) \text{arccot}(u/\bar \lambda).
\end{equation}

In the screeningless limit $\lambda_M/W \rightarrow \infty$ this can be regarded as the derivation of Eq.~\eqref{eq:MirrorVortex}.

\section{Vortex tunneling}
\label{app:VortexTunneling}

In this section we present details on vortex tunneling.

\subsection{Single vortex kink}

We first consider the kink solution, $y_{\rm kink}(\tau)$, for a single vortex tunneling across the system, $y_{\rm kink}(-\mathcal T/2) = -W/2+\xi$ and $y_{\rm kink}(\mathcal T/2) = W/2-\xi$, which solves the equation of motion
\begin{equation}
- m \ddot y + V'(y) = 0,
\end{equation}
where $V(y)$ given by Eq.~\eqref{eq:SingleVortexPot}. 

The tunneling action can be calculated using energy conservation
\begin{eqnarray}
S_{\rm pot} &=& \int_{-W/2+\xi}^{W/2-\xi} dy \sqrt{2m V(y)} \notag \\
&=&  \frac{4W\sqrt{mJ}}{\pi} \left.
\underbrace{\int_{0}^{\pi/2- \bar \xi} d\bar y \sqrt{\alpha/2 + \ln(\cos(\bar y)/\bar{\xi})}}_{I_\alpha(\bar \xi)} \right \vert_{\alpha = 0}
\end{eqnarray}
Here, $\bar \xi = \pi \xi/W$. It is obvious that $I_0(\bar \xi = \pi/2) = 0$, i.e. $f(\pi/2)=0$ as quoted in the main text. 

The tunneling time in the inner part of a wide strip ($W \gg \xi$) can be calculated as
\begin{eqnarray}
\mathcal T &=& \int_{-W/2+\xi}^{W/2-\xi} dy \left .\sqrt{\frac{m}{4 J\left (\frac{\alpha}{2} + \ln \left [\frac{\cos(\bar y)}{\bar \xi}\right ]\right )}} \right \vert_{\alpha = 0}\notag \\
&=& \left .\frac{W}{\pi v} \int_0^{\frac{\pi}{2} - \bar \xi}d \bar y \frac{1}{\sqrt{\frac{\alpha}{2} + \ln \left [\frac{\cos(\bar y)}{\bar \xi}\right ]}} \right \vert_{\alpha = 0}\notag \\
&=& \frac{W}{\pi v} 4\left . \frac{d}{d \alpha} I_\alpha (\bar \xi)\right \vert_{\alpha = 0}.
\end{eqnarray}

Note that there is an additional contribution of order $1/\Delta$ which corresponds to the time to nucleate a vortex and thus is the lower bound for $\mathcal T$ entering $S_{\rm cond} = J \mathcal T$ int he main text. This concludes the derivation of the results presented around Eq.~\eqref{eq:Bounce2D} of the main text.

Using energy conservation one may also evaluate the asympototic trajectory near the turning points, e.g. $y = \mp W/2 \pm \xi \pm \delta y$, where $0<\delta y \ll \xi$ and 
\begin{equation}
 \delta \dot y = 2v\sqrt{\ln( 1+\delta y/\xi )} \simeq 2v \sqrt{{\delta y }/{\xi}},
\end{equation}
which is solved by
\begin{equation}
\delta y(\tau) = v^2 (\tau- \tau_{i,f})^2/\xi, \label{eq:ApproxSol}
\end{equation}
where $\tau_{i,f}$ is the initial or final time.

\subsubsection{Asymptotic evaluation of auxiliary integral}

We here present the asymptotic evaluation of the auxiliary integral $I_\alpha$. Note that the sum $I_0(\bar \xi) + I'_0(\bar \xi)$ is plotted in Fig.~\ref{fig:MultiVortex} c) of the main text. 

\begin{subequations}
\begin{eqnarray}
I_\alpha (x)  &=& \int_x^1 du (- \partial_u \arccos(u)) \sqrt{\frac{\alpha}{2} + \ln(u/x)} \notag \\
&=& \sqrt{\frac{\alpha}{2}} \arccos(x) + \int_x^1 du \frac{\arccos(u) }{2u \sqrt{\frac{\alpha}{2} + \ln(u/x)}} \notag\\
&=& \sqrt{\frac{\alpha}{2}}  \left (\arccos (x)- \left (\frac{\pi}{2} - x\right ) \right ) \notag \\
&+& \left (\frac{\pi}{2}-1\right ) \sqrt{\frac{\alpha}{2} - \ln{x}} + I_{1,\alpha} (x)+ I_{2,\alpha} (x),
\end{eqnarray}
where
\begin{eqnarray}
I_{1,\alpha}(x) &=& \int_x^1 du \sqrt{\frac{\alpha}{2} +\ln[u/x]} \notag \\
&=& \frac{\sqrt{\pi } e^{-\frac{\alpha}{2} } x}{2}  \left[\text{erfi}\left(\sqrt{\frac{\alpha}{2} }\right)-\text{erfi}\left(\sqrt{\frac{\alpha}{2} -\ln (x)}\right)\right] \notag \\
&& + \sqrt{\frac{\alpha}{2} -\ln (x)}-x\sqrt{\frac{\alpha}{2} } \\
I_{2,\alpha}(x) &=& \int_x^1 du\frac{\arccos(u) - \left (\frac{\pi}{2}-u\right )}{2u \sqrt{\frac{\alpha}{2} + \ln(u/x)}} \notag \\
&\simeq& \frac{2 - \pi \ln(2)}{4 \sqrt{\frac{\alpha}{2} - \ln(x)}}.
\end{eqnarray}
\label{eq:InearlyExact}
\end{subequations}

Here, we twice integrated by parts in order to reduce $I_\alpha (x)$ to a simpler integral, $I_{1,\alpha}(x)$, and an integral which is small and determined by $u \sim 1$, $I_{2,\alpha}(x)$, which we evaluated approximately in the limit $x \ll 1$. Here $\text{erfi}$ is the imaginary error function.  Expanding the result up to next to leading order in $1/\sqrt{\frac{\alpha}{2} - \ln(x)}$, we obtain

\begin{equation}
I_\alpha(x) \simeq \frac{\pi}{2}\left [\sqrt{\frac{\alpha}{2} - \ln(x)} - \frac{1}{2} \frac{\ln(2)}{\sqrt{\frac{\alpha}{2} - \ln(x)}} \right ]. \label{eq:IAsymp}
\end{equation}

\subsection{Bounce solution and variational action}
\label{app:Bounce}

We now consider edge unbinding for the potential, Eq.~\eqref{eq:SingleVortexPot} supplemented with a tilt, $V_I = -\beta y$, where $\beta =\Phi_0 I/W$ at small $\beta$.

As an Ansatz, we consider two kinks located at positions $\pm \delta \tau/2$
\begin{equation}
 y_{\rm bounce} = y_{\rm kink}(\tau + \delta \tau/2) - y_{\rm kink}(\tau - \delta \tau/2) - (W/2 -\xi).
\end{equation} 
For $\delta \tau > \mathcal T$, the two kink solutions do not overlap and the action is
\begin{equation}
S_{\rm bounce}(\delta \tau) = 2 S_{\rm kink}^{a)} - \beta (W - 2\xi) \delta \tau.
\end{equation}

However, if $0< (\mathcal T -\delta \tau)/2 \equiv \tau_f \ll \mathcal T$ the overlap leads to non-linear interaction between the kinks. 
Specifically, the $\beta = 0$ part of the potential contribution to the action is
\begin{align}
S_{\beta = 0} 
&\simeq 2\int_{-\infty}^0 d \tau\Big \lbrace \frac{m\dot y_{\rm kink} (\tau + \frac{\delta \tau}{2})^2}{2} + V(y_{\rm kink} (\tau + \frac{\delta \tau}{2})) \notag \\
& +  \delta y_{\rm kink} (\tau) [m \ddot y_{\rm kink} (\tau + \frac{\delta \tau}{2}) - V'(y_{\rm kink} (\tau + \frac{\delta \tau}{2}))] \notag \\
& - \partial_\tau[m \delta y_{\rm kink} (\tau) \dot y_{\rm kink}(\tau + \frac{\delta \tau}{2})] + m \delta \dot y_{\rm kink} (\tau)^2/2\Big \rbrace.
\end{align}
Here, we used the $\tau \rightarrow - \tau$ symmetry of the Ansatz and we have expanded the action in small $\delta y_{\rm kink}(\tau) = y_{\rm kink}(\tau - \delta \tau/2) + (W/2 - \xi)$. The second line vanishes, because $y_{\rm kink}$ solves the equation of motions. This leads to 

\begin{align}
S_{\beta = 0} &= 2 S_{\rm pot} - 2 \int_{0}^{\tau_f} d \tau m\dot y_{\rm kink} (\tau + \frac{\delta \tau}{2})^2 \notag \\
& - 2 m \delta y_{\rm kink} (0) \dot y_{\rm kink} (\delta \tau/2)\notag \\
&+  \int_{- \tau_f}^0 d\tau m \dot y_{\rm kink}( \tau - \frac{\delta \tau}{2})^2 \notag \\
&= 2 S_{\rm kink}^{\rm a)} - \frac{16 m v^4}{3 \xi^2}\tau_f^3.
\end{align}
Here, we have used the approximate solution Eq.~\eqref{eq:ApproxSol} near the turning point. In total this leads to a $\delta \tau$ dependent bounce action

\begin{eqnarray}
S_{\rm bounce}(\delta \tau) &\simeq & 2 S_{\rm kink}^{a)} - \beta (W - 2\xi) \mathcal T \notag \\
&+& \beta (W - 2\xi) (\mathcal T - \delta \tau) \notag \\
&-& E_{\rm cond} (\mathcal T - \delta \tau) \notag\\
&-& \frac{2m v^4}{3 \xi^2} (\mathcal T - \delta \tau)^3
\end{eqnarray}
Here, the second last line is the correction to the condensation energy paid at during the time of the bounce $\mathcal T + \delta \tau$, which is shorter than the time of two kinks. 

This concludes the derivation of Eq.~\eqref{eq:Bounce2D} of the main text.

\section{Multivortex configurations}
\label{app:MultivortexAction}

In this section we present details on the co-tunneling events of multiple vortices.

\begin{figure}
\includegraphics[scale=.7]{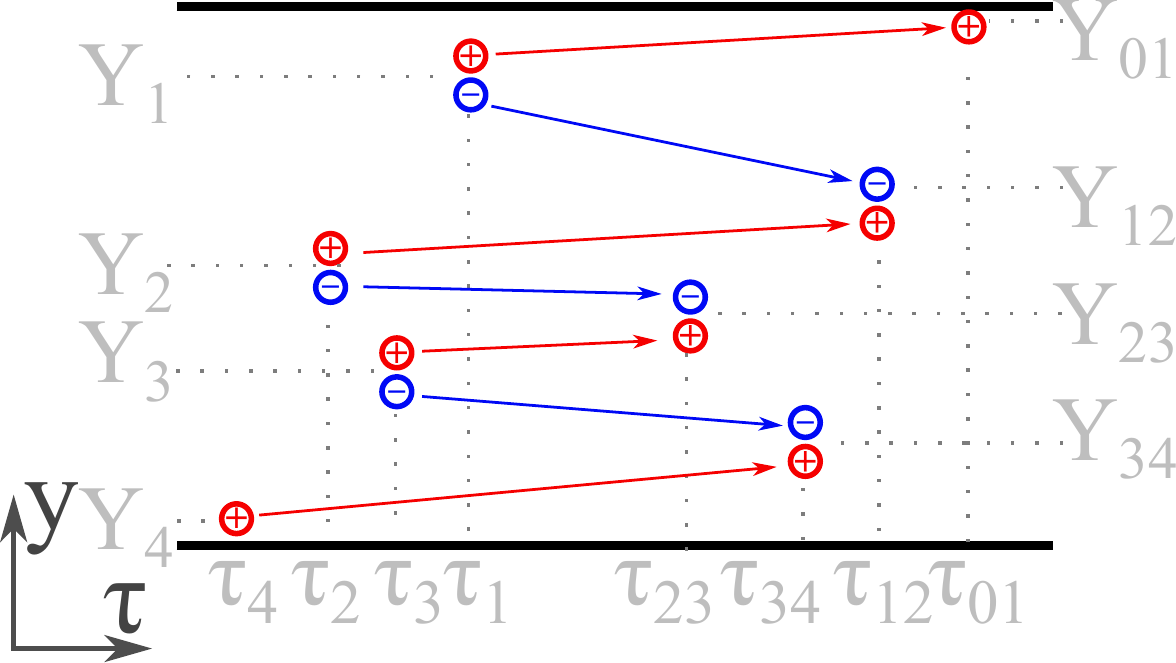} 
\caption{Graphical representation of the variational \textit{Ansatz} \eqref{eq:AnsatzMultivortex} for the action in case a) and $n = 3$. Each arrow contributes $s(d)$, Eq.~\eqref{eq:AnsatzSd}, to the total action, where $d$ is the distance in real space between the points of nucleation and annihilation of the vortices. The major approximation is that the motion along the arrows is linear with constant speed $v(d)$. All space time positions of nucleation except $Y_{n+1} = -W/2$ are variational parameters on which also the space time positions of annihilation depend.}
\label{fig:ProcessAIntegrations}
\end{figure}

\subsection{Variational \textit{Ansatz} and solution}

We consider processes which involve $n$ dipoles in the bulk of the system. We estimate the contribution of these processes by means of the following variational \textit{Ansatz} for the action, see also Fig.~\ref{fig:ProcessAIntegrations}. For the edge unbinding, i.e. processes of type Fig.~\ref{fig:MultiVortex} a) we write
\begin{subequations}
\begin{equation}
S_{a)} = \sum_{i = 1}^{n} S_{\rm kink}(\vert \v X_i - \v X_{i,i+1}\vert) + \sum_{i = 0}^{n}  S_{\rm kink}(\vert  \v X_{i,i+1}- \v X_{i +1} \vert),\label{eq:AnsatzMultivortexa}
\end{equation}
and for the case of Fig.~\ref{fig:MultiVortex} b)
\begin{equation}
S_{b)} = \sum_{i = 1}^{n} S_{\rm kink}(\vert \v X_i - \v X_{i,i+1}\vert) + \sum_{i = 0}^{n-1} S_{\rm kink}(\vert  \v X_{i,i+1}- \v X_{i +1}\vert),\label{eq:AnsatzMultivortexb}
\end{equation}
and we remind the reader that
\begin{equation}
S_{\rm kink}(d) = S_{\rm pot}(d) + S_{\rm cond}(d)\label{eq:AnsatzSd}
\end{equation}
with
\begin{align}
S_{\rm pot}(d) &= 2  m v\, d \sqrt{ - \ln \left ( \frac{2 \xi}{d}\right )}, \\
S_{\rm cond}(d) &=  m v\, d /\sqrt{ - \ln \left ( \frac{2 \xi}{d}\right )},
\end{align}
for $d \gg \xi$, (the general expression is presented in Fig.~\ref{fig:MultiVortex} c).
\label{eq:AnsatzMultivortex}
\end{subequations}
Following the graphical representation of Fig~\ref{fig:MultiVortex} and Fig.~\ref{fig:ProcessAIntegrations}, we label initial positions $\v X_i = (X_i,Y_i)$ and initial times $\tau_i$ with indices $i$ ordered from top to bottom, i.e. $Y_i >Y_{i+1}$. The positions $\v X_{i,i+1} = (X_{i,i+1},Y_{i,i+1})$ and times $\tau_{i,i+1}$ represent the points in space time, at which recombined vortex pairs annihilate. The tunneling event is dominated by contributions which fulfill $Y_{i,i+1} > Y_{i+1,i+2}$. Furthermore, for case a) we define $\v X_{n+1}= (X_{n+1},-W/2)$ and for case b) $\v X_{n,n+1} = (X_{n,n+1},-W/2)$. In both cases a) and b) $\v X_{0,1}= (X_{0,1},W/2)$.

Of course, processes with vortices and antivortices interchanged or processes of the type a) with a net vortex transfer from top to bottom are also present. Their contribution equals the contribution of one of the processes discussed here and are thus not discussed separately.

The calculation involves two assumptions. The first is the approximation of linear motion with constant velocity $v(d)$ in an interval of distance $d$ (e.g. $d = \vert \v X_1 - \v X_{1,2} \vert$). In fact, writing $\mathcal T = d/v(d)$ leads to $v(d)$ which is nearly consistent with the maximal vortex speed as imposed by energy conservation (up to a factor of 2).  The second approximation is that each space time position $(\v X_{i,i+1}, \tau_{i,i+1})$ depends only on the two starting points of adjacent vortices, i.e. on $(\v X_{i}, \tau_{i})$ and $(\v X_{i+1}, \tau_{i+1})$. This second approximation has no influence on the saddle point action, but it does affect the fluctuation determinant. The parameter regime of validity of these approximations shall be discussed below.

\subsubsection{Saddle-point configurations}

We vary the Ansatz~\eqref{eq:AnsatzMultivortex} with respect to the starting times and positions of the vortices. It is convenient to use a lighter notation $S_{\rm kink}(d) = s(d)$. This yields (here $\mu,\nu = x, y$)
\begin{eqnarray}
d_{\v X_{k,\mu}} S_{a),b)} &=& \Big \lbrace s'(\vert \v X_{k-1,k} - \v X_{k} \vert ) \reallywidehat{(\v X_{k-1,k} - \v X_{k} )}_\nu \notag \\
&-&  s'(\vert \v X_{k-1} - \v X_{k-1,k} \vert ) \reallywidehat{( \v X_{k-1} - \v X_{k-1,k})}_\nu  \Big \rbrace \notag \\
&\times& \partial_{\v X_{k,\mu}} (\v X_{k-1,k})_\nu \notag \\
&+&\Big \lbrace s'(\vert \v X_{k,k+1} - \v X_{k+1} \vert ) \reallywidehat{(\v X_{k,k+1} - \v X_{k+1} )}_\nu \notag \\
&-&  s'(\vert \v X_{k} - \v X_{k,k+1} \vert ) \reallywidehat{(  \v X_{k} - \v X_{k,k+1})}_\nu  \Big \rbrace \notag \\
&\times& \partial_{\v X_{k,\mu}} (\v X_{k,k+1})_\nu \notag \\
&-& s'(\vert \v X_{k-1,k} - \v X_{k} \vert ) \reallywidehat{(\v X_{k-1,k} - \v X_{k} )}_\mu \notag \\
&+&  s'(\vert \v X_{k} - \v X_{k,k+1} \vert ) \reallywidehat{(  \v X_{k} - \v X_{k,k+1})}_\mu,
\end{eqnarray}
and
\begin{eqnarray}
d_{\tau_{k}} S_{a),b)} &=& \Big \lbrace s'(\vert \v X_{k-1,k} - \v X_{k} \vert ) \reallywidehat{(\v X_{k-1,k} - \v X_{k} )}_\nu \notag \\
&-&  s'(\vert \v X_{k-1} - \v X_{k-1,k} \vert ) \reallywidehat{( \v X_{k-1} - \v X_{k-1,k})}_\nu  \Big \rbrace \notag \\
&\times& \partial_{\tau_{k}} (\v X_{k-1,k})_\nu \notag \\
&+&\Big \lbrace s'(\vert \v X_{k,k+1} - \v X_{k+1} \vert ) \reallywidehat{(\v X_{k,k+1} - \v X_{k+1} )}_\nu \notag \\
&-&  s'(\vert \v X_{k} - \v X_{k,k+1} \vert ) \reallywidehat{(  \v X_{k} - \v X_{k,k+1})}_\nu  \Big \rbrace \notag \\
&\times& \partial_{\tau_{k}} (\v X_{k,k+1})_\nu.
\end{eqnarray}

Recall that in case a) [case b)] $Y_{01} = W/2$ and $Y_{n+1} = -W/2$ [$Y_{01} = W/2$ and $Y_{n,n+1} = -W/2$] are fixed and thus not variational parameters. By consequence, $\partial_{\v X_1} Y_{01} \equiv 0$ in both cases, and additionally in case b) $\partial_{\v X_n} Y_{n,n+1} \equiv 0$ while in case a) the terms proportional to $\partial_{Y_{n+1}} \v X_{n,n+1}$ should be omitted.

Clearly, a solution of these equations is given by $\reallywidehat{(\v X_{k-1,k} - \v X_{k} )}=\reallywidehat{( \v X_{k-1} - \v X_{k-1,k})} \quad \forall k$, $X_i = X_j \quad \forall i,j$, and equal distances in $y$-direction $Y_{i-1,i}-Y_i = Y_i - Y_{i,i+1} =d_N \quad \forall i$. In case a) $d_N = W/(2n +1)$, and in case b) $d_N = W/(2n)$. By consequence of the linear motion all starting times are the same, $\tau_i = \tau_j \quad \forall i,j$. This is the physical solution we perturb about.

\subsubsection{Determination of space time positions of vortex annihilation}

For the derivation of the fluctuation determinant, it will be necessary to determine the dependence $\v X_{i,i+1} (\v X_i, \tau_i ; \v X_{i+1}, \tau_{i+1})$. Here we present this dependence perturbing weakly about the saddle-point solution.

The \textit{Ansatz} of linear motion implies for the motion of vortices on downward oriented arrows of Fig.~\ref{fig:ProcessAIntegrations}
\begin{subequations}
\begin{equation}
\v x(\tau) = \v X_k - v(\vert \v X_k - \v X_{k,k+1} \vert) \reallywidehat{\v X_k - \v X_{k,k+1}} (\tau - \tau_k)
\end{equation}
while the motion on upwards oriented arrows is 
\begin{equation}
\v x(\tau) = \v X_{k+1} + v(\vert \v X_{k,k+1} - \v X_{k+1} \vert) \reallywidehat{\v X_{k,k+1} - \v X_{k+1}} (\tau - \tau_{k+1})
\end{equation}
\end{subequations}
We determine the moment of annihilation $\tau_{k,k+1}$ by equating the two expressions. This leads to
\begin{equation}
\tau_{k,k+1} = \frac{\Delta X_k   + v(\Delta X_k ) \tau_k + \Delta X_{k+1}  v(\Delta X_{k+1}) \tau_{k+1}}{v(\Delta X_{k}) +  v(\Delta X_{k+1}) },
\end{equation}
where $\Delta X_{k} = \vert \v X_k - \v X_{k,k+1} \vert $ and $\Delta X_{k+1} = \vert \v X_{k,k+1} - \v X_{k+1} \vert$. Inserting this solution into the \textit{Ansatz} for $\v x(\tau)$ in the quasi classical equations of motion yields
\begin{eqnarray}
\v X_{k,k+1} &=& \reallywidehat{\v X_k - \v X_{k + 1}} \frac{v(\Delta X_{k}) v(\Delta X_{k+1})}{v(\Delta X_{k}) +  v(\Delta X_{k+1}) } (\tau_k - \tau_{k +1}) \notag \\
&+&  \frac{ v(\Delta X_{k+1} )  \v X_k +  v(\Delta X_{k}) \v X_{k+1}}{v(\Delta X_{k}) +  v(\Delta X_{k+1}) }.
\end{eqnarray}

\subsubsection{Fluctuation determinant (Gaussian approximation)}

We here present the calculation of the fluctuation determinant. Thus, space time fluctuations of initial positions are treated at Gaussian level.

For the calculation of the fluctuation determinant, we need the derivative of $\v X_{k,k+1}$ with respect to $\v X_{k}$,$\v X_{k+1}$,$\tau_{k}$ and $\tau_{k+1}$ at the saddle point (SP). We thus obtain in the bulk of the system
\begin{subequations}
\begin{eqnarray}
\partial_{\v X_{k,\nu}} (\v X_{k,k+1})_\mu \vert_{SP} &=& \frac{\delta_{\mu \nu}}{2} \notag \\
&=& \partial_{\v X_{k,\nu}} (\v X_{k,k+1})_\mu \vert_{SP} , \\
\partial_{\tau_k} (\v X_{k,k+1})_\mu \vert_{SP}&=& \underbrace{\frac{v(d_N)/2}{1 -  \frac{v'(d_N) d_N}{2v(d_N)} }}_{=: \tilde v_n/2}\delta_{\mu, y} \notag \\
&=& -\partial_{\tau_{k+1}} (\v X_{k,k+1})_\mu \vert_{SP} .
\end{eqnarray}
At the upper boundary we obtain 
\begin{eqnarray}
\partial_{\v X_1,\nu}(\v X_{0,1})_\mu \vert_{SP} &=& \delta_{\mu,\nu} \delta_{\mu,x}, \\
\partial_{\tau_1}(\v X_{0,1})_\mu \vert_{SP} &=& 0.
\end{eqnarray}
an analogous result for the $\partial_{\v X_n,\nu}(\v X_{n,n+1})_\mu \vert_{SP}, \partial_{\tau_n}(\v X_{n,n+1})_\mu \vert_{SP}$ in the case b). For case a), we remind the reader that $Y_{n+1}$ is not a variational parameter.
\end{subequations}

By means of the previous expressions one can determine the correction to the action determining Gaussian fluctuations around the saddle point solution. Direct calculation of second derivatives of the action Eq.~\eqref{eq:AnsatzMultivortex} show that fluctuations in $x$, $y$ and $\tau$ directions do not mix. We obtain
\begin{subequations}
\begin{eqnarray}
\delta S_{a)} &=& \frac{1}{2} \Bigg \lbrace \frac{s'(d_N)}{d_N} \Big [\frac{1}{2} (\delta X_1)^2 - \delta X_1 \delta X_2 + \delta X_2^2 \notag \\
&& - \delta X_2 \delta X_3 + \dots + \delta X_n^2 - \delta X_n \delta X_{n + 1} + \frac{1}{2} \delta X_{n+1}^2 \Big ] \notag \\
&& + s''(d_N) \tilde v_n^2 \Big [\frac{1}{2} (\delta \tau_1)^2 - \delta \tau_1 \delta \tau_2 + \delta \tau_2^2 \notag \\
&& - \delta \tau_2 \delta \tau_3 + \dots + \delta \tau_n^2 - \delta \tau_n \delta \tau_{n + 1} + \frac{1}{2} \delta \tau_{n+1}^2 \Big ] \notag \\ 
&& + s''(d_N)  \Big [\frac{3}{2} (\delta Y_1)^2 - \delta Y_1 \delta Y_2 + \delta Y_2^2 \notag \\
&& - \delta Y_2 \delta Y_3 + \dots - \delta Y_{n-1} \delta Y_{n}+ \delta Y_n^2 \Big ] \Bigg \rbrace 
\end{eqnarray}
and 
\begin{eqnarray}
\delta S_{b)} &=& \frac{1}{2} \Bigg \lbrace \frac{s'(d_N)}{d_N} \Big [\frac{1}{2} (\delta X_1)^2 - \delta X_1 \delta X_2 + \delta X_2^2 \notag \\
&& - \delta X_2 \delta X_3 + \dots + \delta X_{n-1}^2 - \delta X_{n-1} \delta X_{n} + \frac{1}{2} \delta X_{n}^2 \Big ] \notag \\
&& + s''(d_N) \tilde v_n^2 \Big [\frac{1}{2} (\delta \tau_1)^2 - \delta \tau_1 \delta \tau_2 + \delta \tau_2^2 \notag \\
&& - \delta \tau_2 \delta \tau_3 + \dots + \delta \tau_{n-1}^2 - \delta \tau_{n-1} \delta \tau_{n} + \frac{1}{2} \delta \tau_{n}^2 \Big ] \notag \\ 
&& + s''(d_N)  \Big [\frac{3}{2} (\delta Y_1)^2 - \delta Y_1 \delta Y_2 + \delta Y_2^2 \notag \\
&& - \delta Y_2 \delta Y_3 + \dots - \delta Y_{n-1} \delta Y_{n}+ \frac{3}{2}\delta Y_n^2 \Big ] \Bigg \rbrace .
\end{eqnarray}
\label{eq:app:FluctAction}
\end{subequations}
Note that in both cases a) and b) there are two zero modes. They become apparent when we shift the integration variables $\delta X_i \rightarrow \delta X_{i} + \delta {X_{i+1}} \quad \forall i = 1,2,\dots$ (the same for $\tau_i$s), correspond to the center of mass coordinate and are a consequence of translation invariance in $x$- and $\tau$- direction, respectively. Following the standard procedure, zero mode integrals have to be performed exactly and to be kept out of the fluctuation determinant.

The fluctuation determinant is incorporated (up to a constant) into the entropic contribution to the action $\delta S_H$ which in case a) is
\begin{subequations}
\begin{eqnarray}
\delta S_{H, a)} &=& - \ln \Bigg \lbrace \prod_{i = 1}^n \int_{-\infty}^\infty \frac{dY_i}{2\xi} \frac{dX_i}{2\xi} \frac{d \tau_i}{2\xi/v_0} e^{-\delta S_{a)}}\Bigg \rbrace \notag \\
& \simeq & n \ln \left [s''(d_N) (2\xi)^2\right ] + n \ln \left [\tilde v_n/v_0\right ] \notag \\
&&+ \frac{n}{2} \ln \left [(2\xi)^2 s'(d_N)/d_N\right ] 
\end{eqnarray}
while in case b) it is
\begin{eqnarray}
\delta S_{H, b)} &=& - \ln \Bigg \lbrace \prod_{i = 1}^n \int_{-\infty}^\infty \frac{dY_i}{2\xi} \prod_{i = 1}^{n-1} \int_{-\infty}^\infty \frac{dX_i}{2\xi} \frac{d \tau_i}{2\xi/v_0} e^{-\delta S_{b)}}\Bigg \rbrace \notag \\
& \simeq & (n- \frac{1}{2}) \ln \left [s''(d_N) (2\xi)^2\right ] + (n-1) \ln \left [\tilde v_n/v_0\right ] \notag \\
&& + \frac{n-1}{2} \ln \left [(2\xi)^2 s'(d_N)/d_N\right ].
\end{eqnarray}
\end{subequations}

In the concluding section of the main text, we present a simplified estimate, which matches the large $n$ limit of the calculations above, i.e.
\begin{equation}
\delta S_H \sim n \ln [s''(d_N)(2 \xi)^2] = n \ln\left [\frac{\partial^2}{\partial(d_N/2\xi)^2} s(d_N) \right ]
\end{equation}
We can use that $s(d_N)$ is actually a function of $(d_N/2\xi)$, so that modulo unimportant factors of order unity
\begin{equation}
\delta S_H \sim n \ln [s''(d_N)(2 \xi)^2] \sim n \ln\left [s(d_N) \right ],
\end{equation}
where $d_N \sim 2\xi$ at the optimum.

\section{Effect of mesoscopic fluctuations on vortex tunneling}
\label{App:Disorder}

In this section, we estimate the effect of randomness in the superconducting stiffness $J$
(corresponding to the disorder-induced mesoscopic fluctuations of the gap)
on the resistance generated vortex tunneling events.
The effect of gap fluctuations can be modeled by a random stiffness 
$$J(\v x) = J [1 + \rho(\v x) ]$$ 
with 
$$\langle \rho(\v x) \rho(\v x') \rangle = \rho^2 \xi^2 \delta(\v x - \v x'),$$ 
where $ \rho \sim 1/[E_F\tau_{\rm el}(E_F\tau_{\rm el}-\mathcal C)]$ and $\mathcal C \sim 1$
is small for moderate disorder.~\cite{SkvortsovFeigelman2005} 

The tunneling action for a single-vortex tunneling in the presence of such randomness is approximately given by Taylor expansion 
\begin{equation}
S = S_{\rm kink} + \int_{-W/2+\xi}^{W/2-\xi}dy \sqrt{2m V(y)} \rho(\v x)/2, 
\end{equation}
and thus the ``local resistance'' is given by
\begin{equation}
R(x) = R_0 \exp\left [- \int_{-W/2+\xi}^{W/2-\xi} dy \sqrt{2m V(y)} \rho(\v x) \right],
\end{equation}
where $R_0$ is the result without fluctuations.
Then, it readily follows that $R(x)$ is white-noise, log-normal distributed, 
\begin{align}
\langle \ln R(x) \rangle &= \ln R_0, \\
 \langle \ln R(x)\ln R(x') \rangle& =  2m \rho^2
 \xi^2 \delta(x - x') \int_{-W/2+\xi}^{W/2-\xi} dy\,  V(y) .
\end{align}
The total resistance is given by the integral of $R(x)$ over $x$, i.e., by the average local resistance. Both typical and average resistance behave as ${\ln~R\propto-W}$. 

In multi-vortex tunneling events, the tunneling distances $d_i$ depend on the disorder configuration and are generically no longer equal $d_i \neq d$. At the same time, at the level of the approximations involved here, one may estimate the effect of pinning by $d_i = d$
\begin{equation}
R(x) = R_0 \exp\left [- \frac{W}{d}\int_{-d/2+\xi}^{d/2-\xi} dy \sqrt{2m V(y)} \rho(\v x) \right],
\end{equation}
leading to
\begin{align}
\langle \ln R(x) \rangle &= \ln R_0, \\
 \langle \ln R(x)\ln R(x') \rangle &=   
 \frac{2m \rho^2W^2 \xi^2}{d^2}  \delta(x - x') \int_{-\frac{d}{2}+\xi}^{\frac{d}{2}-\xi} dy\,  V(y),
\end{align}
where $d =\mathcal O(\xi)$ for the optimum number of vortices involved in the tunneling event.

We thus find that, both in the single-vortex and multi-vortex tunneling processes, a local resistance $R(x)$ is weakly (for large normal-state conductances) random and log-normal distributed. The exponential (with $W$) scaling of the total resistance of the strip is not affected by fluctuations in this regime.  


\end{document}